\documentclass{article}
\usepackage{amsmath}    

\def\abstract#1{\vskip 7mm 
        \begin{center}{\large Abstract}\par \smallskip
                \begin{minipage}[c]{12cm}
                        \small #1
                \end{minipage}
        \end{center}
}
\def\title#1{\begin{center}{\Large\bf #1}\end{center}}
\def\author#1{\vskip 5mm \begin{center}{#1}\end{center}}
\def\address#1{\begin{center}{\it #1}\end{center}}
\makeatletter
\@ifundefined{lesssim}{}{}
\@ifundefined{gtrsim}{}{}
\def\vereq#1#2{\lower3pt\vbox{\baselineskip1.5pt \lineskip1.5pt
\ialign{$\m@th#1\hfill##\hfil$\crcr#2\crcr\sim\crcr}}}
\makeatother

\begin{document}

\title{%
  Schwarzschild and Friedmann Lemaître Robertson Walker metrics from Newtonian Gravitational collapse
  \smallskip \\
  {\large }
}
\author{%
  Eduardo I. Guendelman\footnote{E-mail:guendel@bgumail.bgu.ac.il},
  Arka Prabha Banik\footnote{E-mail:arkaprab@post.bgu.ac.il},
  Gilad Granit\footnote{E-mail:granitg@post.bgu.ac.il},
  Tomer Ygael\footnote{E-mail:tomeryg@post.bgu.ac.il},
  Christian Rohrhofer\footnote{E-mail:christian.rohrhofer@edu.uni-graz.at}
  }
\address{%
  $^1$ $^2$ $^3$ $^4$Ben Gurion University of the Negev, Department of Physics, Beer-Sheva, Israel\\
  $^5$University of Graz, Institute of Physics, 8010 Graz,Austria
}

\abstract 
{
As  is well known, the $0-0$ component of the Schwarzschild space can be obtained by the requirement that the geodesic of slowly moving particles match the Newtonian equation. Given this result, we shall show here that the remaining components can be obtained by requiring that the inside of a Newtonian ball of dust matched at a free falling radius with the external space determines that space to be Schwarzschild, if no pathology exist. Also we are able to determine that the constant of integration that appears in the Newtonian Cosmology, coincides with the spatial curvature of the FLRW metric.
}

\section{Introduction}
{General Relativity without Einstein's equation}

It is interesting to see how far one can go in the derivation of GR results without using in fact the EE. Sommerfeld \cite{ref1} tried to derive some aspects of General relativity starting from the Schwarschild metric $$ds^{2}=-\gamma^{-2}dt^{2}+\gamma^{2}dr^{2}+r^{2}(d\theta^{2}+sin^{2}\theta d\phi^{2})$$
with $\gamma=1-\frac{2M}{r}$. He then conceived that 

\begin{equation}
dt_{0}=\gamma^{-1}dt,
\end{equation}

\begin{equation}
dr_{0}=\gamma dr,
\end{equation}

He related them with time-dilatation and length-contraction. The time metric and space metric has been shown to be related with opposite power of $\gamma$. But this
construction has been shown to have many flaws \cite{Ref7} \cite{Ref8}.  Instead of pushing ideas along the lines of Sommerfeld, we investigate another direction, based on the fact that in cosmology, it is indeed possible to obtain the Friedmann equations for the expansion
factor $R(t)$. The cosmological solution may cut, thus representing a "ball " of matter, outside this 
spherically symmetric ball we will assume there is a spherically symmetric static space time. Point to be noted that there is no use of Birkhoff's theorem to provide the static assumption. here we have used the analogy of Newtonian gravity or electromagnetism where outside the matter or conducting sphere the field is static. As is well know, any static metric can be brought to the "standard" form, we will be labeling coordinates of the outside space using  bars, so we define  $\bar{t}$ and $\bar{r}$ for the time and radial coordinates outside.
 
\begin{eqnarray}
\label{externalmetric}
ds^2 = - B(\bar{r})d \bar{t}^2 + A(\bar{r})d \bar{r}^2 + \bar{r}^2\left(d\theta^2 + \sin^2 \theta d \phi^2\right).
\end{eqnarray}

From the requirement that Newton's law for gravity holds in some limit, we obtain  an idea of what
 $g_{00} = - B$ 
should be:  indeed, this metric component can be obtained by the requirement that the geodesic of slowly moving particles match the Newtonian equation. For  $g_{\bar{r}\bar{r}} = A(\bar{r})$ is harder.\\
In this paper we develop an approach which will go beyond the simple derivation of the $0-0$ component of the Schwarzschild metric  and give a derivation to obtain the $g_{\bar{r}\bar{r}}$ component under some assumptions that we will specify and avoid previous criticism\cite{Ref2}. The procedure consists of considering a free falling dust ball. As it is known that Newtonian cosmology \cite{Ref4} can be
developed and a homogeneous dust system can be studied and by applying the Newton’s laws.We can obtain that each dust particle expands or contracts according to a single expansion factor $R$, obeying
\begin{eqnarray}
\left( \frac{\dot{R}}{R} \right)^2 = \frac{8 \pi G}{3} \rho - \frac{k}{R^2},
\end{eqnarray}
where, $k$ being an integration constant. This is possible for considering zero Cosmological constant. Points to be noted that this equation came out from the second order differential equation of R.

\medskip

So, the radius for a given particle is proportional to $R(t)$, By using the continuity equation and the Euler motion equation, we can obtain that the density $\rho \sim \frac{\rho_0}{R^3}$ and if we require $R(0) = 1, \dot{R}(0) = 0 , \rho = \rho(0)$ at $t=0$, we get
\begin{eqnarray}
\label{the value of k}
k = \frac{8 \pi G}{3} \rho(0) > 0 \quad \textsl{and} \quad \dot{R}(t)^2 = k \left[\frac{1}{R} - 1 \right].
\end{eqnarray}
In the present work we will not require or derive or motivate a geometric interpretation of the constant of integration $k$. In contrast, a follow up section will see that, even with out using EE, $k$ can be given the interpretation of spatial curvature in FRW space. Then matching to an external space can be used to determine the outside space by demanding continuity in some coordinate system from the inside to the outside. Of course, the face one can give a geometrical interpretation to the constant $k$ without using EE is by itself interesting. Here we only will assume that the metric inside the ball is of the form
\begin{eqnarray}
\label{dt2}
ds^2 = -dt^2 + g_{ij} dx^i dx^j
\end{eqnarray}
and that the dust particles are co-moving that is $dx^i=0$ which are geodesics of the space time\cite{Ref3}. Furthermore, just from homogeneity and isotropy, we can determine that 
\begin{eqnarray}
\label{gij}
g_{ij} dx^i dx^j = R^2(t)\left(\frac{dr^2}{1-\kappa r^2} + r^2\left(d\theta^2 + \sin^2\theta d\phi\right),\right)
\end{eqnarray}
where the radial and time coordinates are $r$ and $t$ respectively. Notice that apriori we do not know if $k$ and $\kappa$ are related.
A very basic and elementary relation that we obtain when matching the space time (\ref{externalmetric})
with (\ref{dt2} - \ref{gij}), is obtained when considering the length  of a circle of an equatorial circle
($sin\theta =1$)as $\phi$ runs from $0$ to $2\pi$. The inside observer (in the dust ball), just below the surface of matching will measure a length $2\pi r R$, while The outside observer (in the static space), just above the surface of matching will measure a length $2\pi \bar{r}$, since these two measurements refer to the same physical length, in order for the complete geometry to be well defined, we obtain
\begin{eqnarray}
\label{bar}
\bar{r} = r R(t).
\end{eqnarray}

\section{Newtonian Cosmology}
Following the Cosmological Principle, in a homogeneous and isotropic universe a spherical region of radius $R$ can be identified around an arbitrary point, whose matter density $\rho$ within is homogeneous. The surrounding matter cannot have any influence on its dynamics, as this would violate isotropy. By this general assumptions, the size of the sphere is arbitrary.
The equation of motion for a test mass $ m $ located on the boundary of such a sphere shall be described in terms of a homogeneous positive parameter $R(t)$, where the coordinate of each particle expands according to $a(t)=constant \cdot R(t)$, where the constant depends on the particular particle. Along with the assumption of the motion of the particles as follows Bondi's book \cite{Ref4}.

 \begin{equation}
\overrightarrow{v}=f(t)\overrightarrow{a},
\end{equation}

\begin{equation}
\rho=\rho(t),
\end{equation}

\begin{equation}
P=P(t)=0,
\end{equation}

here we are discussing the situation with $P=0$.

\begin{equation}
\frac{1}{R}\frac{dR}{dt}=f(t)\;,R(t=0)=1,
\end{equation}

Now by applying Euler's equation of motions , we have 

\begin{equation}
\frac{D\overrightarrow{v}}{Dt}-\overrightarrow{F}=0=\overrightarrow{r}[\frac{df}{dt}+f^{2}]-\overrightarrow{F},
\end{equation}

where the Hydrodynamic operator is $\frac{D}{Dt}=\frac{\partial}{\partial t}+(\vec{v.}\nabla)$,

 now, taking divergence on both sides 

\begin{equation}
3[\frac{df}{dt}+f^{2}]=\nabla.\overrightarrow{F}
\end{equation}

where, $\overrightarrow{F}$ is the body force per unit mass.

So by Poisson's equation 

\begin{equation}
\nabla.\overrightarrow{F}=-4\pi G\rho
\end{equation}

So, we have then

\begin{equation}
3[\frac{df}{dt}+f^{2}]=-4\pi G\rho
\end{equation}

Putting the value of $f(t)$ in the previous equation imply a similar equation for the universal expansion factor $R(t)$

\begin{equation}
\label{eq:newton1}
\ddot{R}=-\frac{G}{R^2} \left(\frac{4\pi}{3}R^3 \rho \right)= - \frac{4\pi G}{3} R\rho.
\end{equation}

Basically, this corresponds to Friedmann's second equation without a cosmological constant $\Lambda$ and zero pressure.
As the linear dimensions scale by $R(t)$, all co-moving volumes should scale by $R(t)^3$, that is a $1/R^{3}$ dependency for the density, which dilutes the matter as the sphere expands.

For deriving a second equation, we first consider mass conservation within co-moving sphere,
\begin{equation}
\label{eq:energyconservation1}
\frac{d}{dt}\left(\frac{4\pi}{3}  R^{3}\rho  \right) = 0,
\end{equation}
where the internal mass $M$ inside the sphere should be constant . By performing the derivative and simplifying one $R$, the equation gets
\begin{equation}
2R\dot{R}\rho + R\dot{R}\rho + R^2 \dot{\rho} = 0.
\end{equation}
The second term can be eliminated by (\ref{eq:newton1}) and after restoring derivatives the equation
\begin{equation}
\frac{d(\dot{R}^2)}{dt} = \frac{8\pi G}{3} \frac{d(\rho R^2)}{dt}
\end{equation}
is obtained. Integration on both sides gives
\begin{equation}
\dot{R}^2 = \frac{8\pi G}{3} \rho R^2 - \tilde{k},
\end{equation}
and rewriting the arbitrary integration constant $\tilde{k}$ in a way to match the units $\tilde{k} \longrightarrow kc^2$ yields finally an equation, which corresponds to Friedmann's first equation in structure:  
\begin{equation}
\label{eq:fridman1}
\left(\frac{\dot{R}}{R}\right)^{2} = \frac{8\pi G}{3} \rho - k \left(\frac{c}{R}\right)^2.
\end{equation}
It is interesting to point out the physical meaning of the constant $k$. Indeed, it is very much like a conserved energy of a mechanical system, and indeed the sign of $k$ determines whether $R$ will expand to infinity, or expand, achieve a maximum and then re collapse. Indeed, a simple analysis shows that for 
$k>0$, we have a bound system, a bound system that has an associated negative energy, like wise $k<0$
represents a case where the expansion factor reaches infinity and is the analogous of a mechanical system 
with positive energy. Finally, a case where $R$ barely makes it to infinity is $k=0$, and corresponds to a system with zero energy, analogous to the very similar problem of the critical escape trajectory that just makes in order to escape the earth.
A separate issue is a possible geometrical interpretation of $k$, in particular a a possible relation between $k$ and the $\kappa$ that appears in (\ref{gij}), this we will achieve, without assuming Einstein's equations.
The FLRW and the appearance of $R(t)$ in (\ref{gij}) can be justified on purely symmetry and geometrical grounds, independent of any specific dynamics, General Relativity or otherwise. 
As we pointed out, although we will not invoke Einstein's equations, we will make use of the much simpler geodesic equation. As we mentioned, it is a rather simple matter to show that "co-moving observers",
of the metric (\ref{dt2}), (\ref{gij}), that is those trajectories where the spatial coordinates in 
(\ref{dt2}),(\ref{gij}) are constants are indeed geodesics. Since, $x^{i}=const$ is a geodesic in FLRW spacetime as FLRW metric yields $\Gamma_{tt}^{\mu}=0$ \cite{Weinberg} . Physical distances between these  "co-moving observers" scale as $R$, so the $R$ in the Newtonian cosmology coincides with the $R$ that appears in (\ref{gij}), the relation between
$k$ and $\kappa$ , this will require a bit more elaboration (without using EE).
We can now describe cosmic large scale dynamics by a single expansion parameter $R(t)$, as long the cosmic principle holds, and the predictions corresponds to GR.
Note that from here on we will be using $c=1$.

\section{Finding $g_{00}$}

The $g_{00}$ element of the metric is the first thing that students start form when learning GR, which a student might find hard to grasp. This is most likely the result of still not willing to be parted from Newtonian ideas, while accepting the ideas of special relativity.

In this section, in order not to make the notation to heavy, we will for the moment denote the
coordinates associated with the static space time simply t, r, $\theta$ and $\phi$. In the next section, we go back to the bar notation when it refers to an external space and unbarred coordinates for the internal cosmological space time .
In the next two subsections we will derive the $g_{tt}$ element of the metric from the geodesic equation in two different methods, which are really equivalent, but we want to present both to give the possible instructor some alternatives, or to present both.

We allow our usage of the geodesic equation, because this subject does not involve a high level of complexity as compared to the development of the Einstein's equations, and it can be derived from the  from the principle of least action of the particle trajectory, the action being the proper time along the trajectory of the particle in a certain given metric. Starting from the interval
\begin{eqnarray}
 ds^{2}=g_{\mu\nu}dx^{\mu}dx^{\nu} = -d\tau^2,
\end{eqnarray}
we integrate over $d\tau$ and by using a monotonic variation parameter $\sigma$ on the particle trajectory
\begin{eqnarray}
\mathcal{\tau}=\int\mathbf{d}\sigma\sqrt{-g_{\mu\nu}\frac{dx^{\mu}}{d\sigma}\frac{dx^{\nu}}{d\sigma}},
\end{eqnarray}
from here applying the principle of least action we find the geodesic equation
\begin{eqnarray}
 \frac{d^{2}x^{\mu}}{d\tau^{2}} + \Gamma_{\alpha\beta}^{\mu}\frac{dx^{\alpha}}{d\tau}\frac{dx^{\beta}}{d\tau} = 0,
\end{eqnarray}
with $\Gamma_{\alpha\beta}^{\mu}$ known as the Christoffel symbol (or connection)
\begin{eqnarray}
 \Gamma_{\alpha\beta}^{\mu}=\frac{1}{2}g^{\mu\rho}\left(\frac{\partial g_{\rho\beta}}{\partial x^{\alpha}}+\frac{\partial g_{\alpha\rho}}{\partial x^{\beta}}-\frac{\partial g_{\alpha\beta}}{\partial x^{\rho}}\right).
\end{eqnarray}
Remember that $\Gamma_{\alpha\beta}^{\mu}$ is not a tensor, but the sum of the two terms in the geodesic equation is a vector.
\\
 
\textbf{1st Method: Non - Relativistic limit}\\
The following method starts by a simple exchange of derivatives $\frac{d}{dt }=\left(\frac{dt}{d\tau}\right)^{-1}\frac{d}{d\tau}$, plugging in to the $x^{i}$ part of the equation LHS, we have :
\begin{eqnarray}
\frac{d^{2}x^{i}}{dt^{2}}=\left(\frac{dt}{d\tau}\right)^{-1}\frac{d}{d\tau}\left(\left(\frac{dt}{d\tau}\right)^{-1}\frac{dx^{i}}{d\tau}\right)=\left(\frac{dt}{d\tau}\right)^{-2}\frac{d^{2}x^{i}}{d\tau^{2}}-\left(\frac{dt}{d\tau}\right)^{-3}\frac{d^{2}t}{d\tau^{2}}\frac{dx^{i}}{d\tau}.
\end{eqnarray}
Now using the geodesic equation, the RHS can be re written as,
\begin{eqnarray}
-\left(\frac{dt}{d\tau}\right)^{-2}\Gamma_{\alpha\beta}^{i}\frac{dx^{\alpha}}{dt}\frac{dx^{\beta}}{dt}+\left(\frac{dt}{d\tau}\right)^{-3}\Gamma_{\alpha\beta}^{0}\frac{dx^{\alpha}}{dt}\frac{dx^{\beta}}{dt}\frac{dx^{i}}{dt},
\end{eqnarray}
now equating both sides
\begin{eqnarray}
 \frac{d^{2}x^{i}}{dt^{2}}=-\Gamma_{\alpha\beta}^{i}\frac{dx^{\alpha}}{dt}\frac{dx^{\beta}}{dt}+\Gamma_{\alpha\beta}^{0}\frac{dx^{\alpha}}{dt}\frac{dx^{\beta}}{dt}\frac{dx^{i}}{dt} .
\end{eqnarray}
We continue this derivation by ordering in terms of $ v^{i}=\tfrac{dx^{i}}{dt}$, by taking the following limit $v^{2}\sim\frac{GM}{\bar{r}}$ and $\frac{dt}{d\tau}=\mathcal{O}\left(1\right)$

\begin{eqnarray}
  \frac{d^{2}x^{i}}{dt^{2}}=-\Gamma_{tt}^{i}-2\Gamma_{tj}^{i}\frac{dx^{j}}{dt}-\Gamma_{jk}^{i}\frac{dx^{j}}{dt}\frac{dx^{k}}{dt}+\left(\Gamma_{tt}^{t}+2\Gamma_{tj}^{t}\frac{dx^{j}}{dt}+\Gamma_{jk}^{t}\frac{dx^{j}}{dt}\frac{dx^{k}}{dt}\right)\frac{dx^{i}}{dt} ,
\end{eqnarray}
to derive the Newtonian limit we only consider the first term, since this is the term that dominates in the limit of very small velocities, 

\begin{eqnarray}
\frac{d^{2}x^{i}}{dt^{2}}=-\Gamma_{tt}^{i}=-\frac{1}{2}g^{i\rho}\left(\frac{\partial g_{\rho t}}{\partial x^{t}}+\frac{\partial g_{t\rho}}{\partial x^{t}}-\frac{\partial g_{tt}}{\partial x^{\rho}}\right)=\frac{1}{2}g^{ij}\frac{\partial g_{tt}}{\partial x^{j}} \approx \frac{1}{2}\delta^{ij}\frac{\partial g_{tt}}{\partial x^{j}}.
\end{eqnarray}
In the above we have used that $g^{ij}$ is in zero-th order approximation equal to $\delta^{ij}$, since the deviations from Minkowski space-time should be indeed very small.
By demanding even for the small deviations to vanish at infinity, i.e. that asymptotic limit of the metric is exactly Minkowski space-time $\underset{r\rightarrow\infty}{\lim}g_{tt}=-1$ then by comparing with the Newtonian force law of universal gravity,
\begin{eqnarray}
\frac{d^2 x}{dt^2} = - \nabla \phi \quad \textsl{where} \quad \phi = -\frac{GM}{r},
\end{eqnarray}
we simply solve
\begin{eqnarray}
g_{tt}=-\left(1-\frac{2GM}{r}\right)=-\left(1+2\phi\right).
\end{eqnarray}\\

\textbf{2nd Method: Metric perturbation}\\
The following method starts with taking two assumptions :
\begin{enumerate}
\item  The metric is static with a symmetric perturbation 
\begin{eqnarray}
g_{\mu\nu}=\eta_{\mu\nu}+h_{\mu\nu}\;;\;\left|h_{\mu\nu}\right|\ll1 \quad \textsl{and} \quad g_{\mu\nu}\approx\eta_{\mu\nu}\;;\; g^{\mu\nu}\approx\eta^{\mu\nu},
\end{eqnarray}
\item Non-relativistic limit
\begin{eqnarray}
v^{i}=\frac{dx^{i}}{dt}\ll1\rightarrow\frac{dx^{i}}{d\tau}\ll\frac{dt}{d\tau}\sim1,
\end{eqnarray}
\end{enumerate}
We now return to the geodesic equation
\begin{eqnarray}
 \frac{d^{2}x^{\mu}}{d\tau^{2}}=-\Gamma_{\alpha\beta}^{\mu}\frac{dx^{\alpha}}{d\tau}\frac{dx^{\beta}}{d\tau}=-\Gamma_{tt}^{\mu}\frac{dx^{t}}{d\tau}\frac{dx^{t}}{d\tau}-\Gamma_{ij}^{\mu}\frac{dx^{i}}{d\tau}\frac{dx^{j}}{d\tau}\approx-\Gamma_{tt}^{\mu},
\end{eqnarray}
writing the equations for the spatial part
\begin{eqnarray}
 \Gamma_{tt}^{i}=-\frac{1}{2}\eta^{ij}\left(\frac{\partial g_{tt}}{\partial x^{j}}\right)=\frac{1}{2}\delta^{ij}\left(\frac{\partial g_{tt}}{\partial x^{j}}\right)
 \frac{d^{2}x^{i}}{dt^{2}}=\frac{1}{2}\left(\frac{\partial g_{tt}}{\partial x^{i}}\right).
\end{eqnarray}
Now, by demanding the asymptotic limit $\underset{r\rightarrow\infty}{\lim}g_{tt}=-1$, we once again comparing ($21$) to Newtonian force law of universal gravity as in the previous sub-section
\begin{eqnarray}
 g_{tt}=-\left(1-\frac{2GM}{r}\right)=-\left(1+2\phi\right).
\end{eqnarray}

As we mentioned before, the two methods are equivalent, but the first method gives us some idea of what terms we are dropping when doing our approximations, while the second method does not.

\section{Finding $g_{\bar{r}\bar{r}}$}
The  $g_{\bar{r}\bar{r}}$ of the metric in the outside static region in our case,  has been the "elusive component", this component has not been calculated using matching to a Newtonian cosmology previously, here we will show that this is possible. In this section we will find $g_{\bar{r}\bar{r}}$ using the assumption that we have a co-moving observer satisfying $r = constant$. We will also assume that inside and at the boundary of the dust ball the radius evolves as $\bar{r}= r R(t)$, where $R(t)$ is determined by (\ref{eq:fridman1}). We also assume that for $\bar{r} > r R(t)$ the motion is on a radial geodesic and the metric is of the form
\begin{eqnarray}
ds^2 &=& - \left( 1 - \frac{2GM}{\bar{r}}\right) d\bar{t}^2 + A(\bar{r}) d\bar{r}^2 + \bar{r}^2 d \Omega^2 \\
d \Omega^2 &=& d \theta^2 + \sin^2 \theta d \phi^2 ,
\end{eqnarray}
we labeled the time as $\bar{t}$ because it may not be the same coordinate as the $t$ in the dust ball and the  $g_{tt}$ component was found in the previous section. Notice that we assume that the external metric is time independent.\\
A radially falling geodesic, meaning that $\theta = const$ and $\phi = const$, is fully described by the conservation of energy that results from that the metric outside is assumed to be static.
The geodesics are derived from the action 
\begin{eqnarray}
S = \int d\sigma \sqrt{-\frac{d x^\mu}{d \sigma} \frac{d x^\nu}{d \sigma} g_{\mu \nu}(x)}.
\end{eqnarray}
The equation with respect to $\bar{t}$ is
\begin{equation}
\frac{d}{d\sigma}\left( \frac{\partial L}{\partial \dot{\bar{t}}}\right) = 0 \quad \textsl{where} \quad \dot{\bar{t}} = \frac{d \bar{t}}{d \sigma}.
\end{equation}

This gives us
\begin{equation}
\gamma = \frac{\partial L}{\partial \dot{\bar{t}}} = \left( 1 - \frac{2GM}{\bar{r}} \right) \frac{d \bar{t}}{d \tau},
\end{equation}
where $\gamma$ is constant and $d\tau$ is the proper time.\\
Notice that since the spatial coordinates in the space $d x^i = 0$ we get 
\begin{equation}
d\tau = d t ,
\end{equation}
giving us
\begin{eqnarray}
d \tau^2 &=&  \left( 1 - \frac{2GM}{\bar{r}} \right) d \bar{t}^2 - A(\bar{r})d\bar{r}^2\\
\left( \frac{d \tau}{d t} \right)^2 = 1 &=& \left( 1 - \frac{2GM}{\bar{r}} \right) \left( \frac{d \bar{t}}{dt} \right)^2 - A(\bar{r}) \left( \frac{d\bar{r}}{dt} \right)^2 .
\end{eqnarray}
Using previous equations, we obtain
\begin{eqnarray}
\left( 1- \frac{2GM}{\bar{r}}\right)^{-1}\gamma^2 - A(\bar{r}) \left( \frac{d\bar{r}}{dt} \right)^2 = 1 .
\end{eqnarray}
As we have seen, the consistency of the matching of the two spaces requires
 $\bar{r} = R(t) r$, furthermore, we assume that even the boundary of the dust shell
 free falls according to a co-moving observer, which means that the FLRW coordinate
 $r= constant$ and this allow then to solve for  $A(\bar{r})$, 
\begin{eqnarray}
A(\bar{r}) = - \left( 1 - \frac{\gamma^2}{\left( 1 - \frac{2GM}{\bar{r}} \right)}\right)\frac{1}{r^2 \left( \frac{d R}{dt}\right)^2}.
\end{eqnarray}
We now use $(3)$ and get

\begin{eqnarray}
A(\bar{r}) = -\left( 1 - \frac{\gamma^2}{\left( 1 - \frac{2GM}{\bar{r}} \right)}\right)\frac{1}{r^2 k \left( \frac{1}{R} -1 \right)},
\end{eqnarray}
by simplifying this and expressing in terms of $\bar{r}$ we get
\begin{eqnarray}
A(\bar{r}) = \frac{1}{r^2}\frac{\gamma^2 - 1 + \frac{2GM}{\bar{r}}}{1 - \frac{2GM}{\bar{r}}}\frac{1}{k \left( -1 + \frac{r}{\bar{r}}\right)}.
\end{eqnarray}

If we take the limit $\bar{r} \rightarrow \infty $, we see that 
$A(\bar{r}) \rightarrow -(\gamma^2 - 1)/k r^2 $. Asymptotic flatness would require 
$A(\bar{r}) \rightarrow 1$. 

The metric component $A(r)$ is free of singularities (real singularities, not coordinate singularities) and preserves its signature (one time and three spaces)and is asymptotically flat only if
\begin{eqnarray}
\label{A}
\gamma^2 - 1 = - k r^2 \quad \textsl{and} \quad \frac{2GM}{\bar{r}} = \frac{r^2 k}{R} = \frac{k r^3}{\bar{r}} \Rightarrow k = \frac{2GM}{r^3}.
\end{eqnarray}
As mentioned earlier, $r$ is the comoving coordinate and hence constant, whereas $R$ is variable.

\

Notice that above the condition,(\ref{A}) $ k = \frac{2GM}{r^3}$ , when combined with the value of k, as given by
(\ref{the value of k}), $k = \frac{8 \pi G}{3} \rho(0)$ yields us the value of M as
\begin{eqnarray}
M=\frac{4}{3}\pi\rho_{0}r^{3}.
\end{eqnarray}

Surprisingly, it looks like the $Mass= density \times Vol. of E^{3} ball. $ But, we got this relation in consequences of our previous derivation and the flat space volume of the ball has not been used anywhere for deriving this. 
\
\medskip

Finally, all of this gives us
\begin{eqnarray}
A(\bar{r}) = \frac{1}{1 - \frac{2GM}{\bar{r}}},
\end{eqnarray}
reproducing the Schwarzschild spacetime.

\section{Finding the geometric interpretation of the Newtonian Cosmology}
The main aim of this section is to show that for the geometric parameter defining the FRW space coincides with the Newtonian Energy in $k$ found from the interpretation of the Newtonian cosmology and to demonstrate that $k=\kappa$.
\medskip
In an embedding four dimensional Euclidean Space with metric
\[
dl^2 = dx^{2}+dy^{2}+dz^{2}+dw^{2},
\]
we define a 3-sphere if the sets of points ($x,y,z,w$) satisfy
the constraint
\[
x^{2}+y^{2}+z^{2}+w^{2}=1/ \kappa\: with\: \kappa >0,
\]
then it is said to form a three - sphere $S^{3}$.
That is, we will take our cosmological solution to have positive spatial curvature $\kappa>0$ therefore topologically 
spatial slices have the topology of $S^{3}$.
Solving for $w$ from the above constraint and inserting into the expression for $dl^2$  
defining $r^2 = x^{2}+y^{2}+z^{2}$,  $x=rsin\theta cos \phi$
and $y=rsin\theta sin\phi$,$z=rcos\theta$, we obtain the metric of the $3$-sphere,
\[
ds_{3}^{2}=[\frac{dr^{2}}{1-\kappa r^{2}}+r^{2}d\Omega^{2}],
\]
with 
\[
d\Omega^{2}=(d\text{\ensuremath{\theta}}^{2}+sin^{2}\theta d\phi^{2}),
\]
to obtain the physical space time, that is the line element in FLRW Space, we add time and a scale factor that multiplies the $3$-sphere, 
we obtain,
\[
ds^{2}=-dt^{2}+R^{2}(t).ds_{3}^{2} .
\]
So our infinitesimal line element in
FRW space becomes, 
\begin{eqnarray}
ds^{2}=-dt^{2}+R(t)^{2}\left(\frac{dr^{2}}{1-\kappa r^{2}}+r^{2}d\Omega^{2}\right).
\end{eqnarray}
We assume again that that we have a co-moving observer which satisfies $r = const$. Independently of that,
in the FLRW space we can use everywhere (not just at the boundary) the barred radius  $\bar{r}=R(t)r$, which means $r=\frac{\bar{r}}{R(t)}$.\\
This yields
\begin{eqnarray}
dr=\frac{d\bar{r}}{R(t)}-\frac{\dot{R\bar{r}}}{R^{2}}dt.
\end{eqnarray}
Putting this in $(35)$, we've
\begin{eqnarray}
ds^{2}=\left(-1+\frac{\dot{R}^{2}\bar{r}}{(1-\kappa r^{2})R^{2}}\right)dt^{2}+\frac{d\bar{r}^{2}}{1-\kappa r^{2}}-\frac{2\dot{R}\bar{r}}{(1-\kappa r^{2})R}dtd\bar{r}.
\end{eqnarray}
Now, we make a transformation $t=t(\bar{t},\bar{r})$, so infinitesimal change in time 
\begin{eqnarray}
dt=\frac{\partial t}{\partial\bar{r}}d\bar{r}+\frac{\partial t}{\partial\bar{t}}d\bar{t},
\end{eqnarray}
squaring, 
\begin{eqnarray}
dt^{2}=\left(\frac{\partial t}{\partial\bar{r}}\right)^{2}d\bar{r}^{2}+\left(\frac{\partial t}{\partial\bar{t}}\right)^{2}d\bar{t}^{2}+2\left(\frac{\partial t}{\partial\bar{r}}\right)\left(\frac{\partial t}{\partial\bar{t}}\right)d\bar{t}d\bar{r}.
\end{eqnarray}
Under these circumstances, the infinitesimal line element becomes

\begin{eqnarray}
ds^{2}=\left(-1+\frac{\dot{R}^{2}\bar{r}}{(1-\kappa r^{2})R^{2}}\right)\left(\left(\frac{\partial t}{\partial\bar{r}}\right)^{2}d\bar{r}^{2}+\left(\frac{\partial t}{\partial\bar{t}}\right)^{2}d\bar{t}^{2}+2\left(\frac{\partial t}{\partial\bar{r}}\right)\left(\frac{\partial t}{\partial\bar{t}}\right)d\bar{t}d\bar{r}\right)\\
+\frac{d\bar{r}^{2}}{1-\kappa r^{2}}-\frac{2\dot{R}\bar{r}}{\left(1-\kappa r^{2}\right)R}d\bar{r\left(\frac{\partial t}{\partial\bar{r}}d\bar{r}+\frac{\partial t}{\partial\bar{t}}d\bar{t}\right)}.
\end{eqnarray}
Now, we have to eliminate the cross terms, so
\begin{eqnarray}
-\frac{2\dot{R}\bar{r}}{(1-\kappa r^{2})R}d\bar{r\frac{\partial t}{\partial\bar{t}}d\bar{t}}+\left(-1+\frac{\dot{R}^{2}\bar{r}}{(1-\kappa r^{2})R^{2}}\right)2\left(\frac{\partial t}{\partial\bar{r}}\right)\left(\frac{\partial t}{\partial\bar{t}}\right)d\bar{t}d\bar{r}=0 .
\end{eqnarray}
This gives
\begin{eqnarray}
\left(\frac{\partial t}{\partial\bar{r}}\right)&=&\frac{\bar{r}\dot{R}}{(1-\kappa r^{2})R\left(-1+\frac{\dot{R}^{2}\bar{r}}{(1-\kappa r^{2})R^{2}}\right)}\\
&=&\frac{\bar{r}\dot{R}}{-(1-\kappa r^{2})R+\frac{\dot{R}^{2}\bar{r^{2}}}{R}}\\
&=&\frac{\bar{r}\dot{RR}}{-(1-\kappa r^{2})R^{2}+\dot{R}^{2}\bar{r}^{2}}.
\end{eqnarray}
Now lets linger on $g_{\bar{r}\bar{r}}$ with this new $(\frac{\partial t}{\partial\bar{r}})$ value
\begin{gather*}
g_{\bar{r}\bar{r}}=\frac{1}{1-\kappa r^{2}}+\left(-1+\frac{\dot{R}^{2}\bar{r}^{2}}{(1-\kappa r^{2})R^{2}}\right)(\frac{\partial t}{\partial\bar{r}})^{2}-\frac{2\dot{R}\bar{r}}{(1-\kappa r^{2})R}\frac{\partial t}{\partial\bar{r}}\\
\qquad=\frac{1}{1-\kappa r^{2}}+\left(-1+\frac{\dot{R}^{2}\bar{r}^{2}}{(1-\kappa r^{2})R^{2}}\right)\left(\frac{\partial t}{\partial\bar{r}}\right)^{2}-\frac{2\dot{R}\bar{r}}{(1-\kappa r^{2})R}\frac{\partial t}{\partial\bar{r}}\\\nonumber 
\quad\qquad\qquad\qquad=\frac{1}{1-\kappa r^{2}}+\left(-1+\frac{\dot{R}^{2}\bar{r}^{2}}{(1-\kappa r^{2})R^{2}}\right)\left(\frac{\dot{R}^{2}\bar{\mbox{r}^{2}}}{R^{2}(1-\kappa r^{2})\left(-1+\frac{\dot{R}^{2}\bar{r}^{2}}{(1-\kappa r^{2})R^{2}}\right)}\right) -\\ 
-\frac{2\dot{R}\bar{r}}{(1-\kappa r^{2})R}\frac{\dot{R\bar{r}}}{R(1-\kappa r^{2})\left(-1+\frac{\dot{R}^{2}\bar{r}^{2}}{(1-\kappa r^{2})R^{2}}\right)}\\
\qquad\qquad=\frac{1}{1-\kappa r^{2}}\left(1+\frac{\dot{R}^{2}\bar{r}^{2}}{-R^{2}+\kappa R^{2}r^{2}+\dot{R}^{2}\bar{r}^{2}}-\frac{2\dot{R}^{2}\bar{r}^{2}}{-R^{2}+\kappa R^{2}r^{2}+\dot{R}^{2}\bar{r}^{2}}\right)\\
=-\frac{R^{2}(1-\kappa r^{2})}{(1-\kappa r^{2})\left(-R^{2}+\kappa R^{2}r^{2}+\dot{R}^{2}\bar{r}^{2}\right)}\\
=\frac{1}{1-\kappa r^{2}-\frac{\dot{R}^{2}\bar{r}^{2}}{R^{2}}}.
\end{gather*}
For $\bar{r}=R(t)r$ it yields
\begin{eqnarray}
g_{\bar{r}\bar{r}}=\frac{1}{1-r^{2}(\kappa+\dot{R^{2}})} .
\end{eqnarray}
In the last section we derived also what this metric component should be,  
\begin{eqnarray}
g_{\bar{r}\bar{r}}=\frac{1}{1-\frac{2GM}{\bar{r}}}.
\end{eqnarray}
Since from Newtonian cosmology , we know that 
\begin{eqnarray}
k+\dot{R}^{2}=\frac{8\pi}{3R}G\rho_{0}=\frac{8\pi r^{3}\rho_{0}}{3r^{2}(rR)}.
\end{eqnarray}
Therefore,
\begin{eqnarray}
(k+\dot{R}^2)r^{2}=\frac{8\pi r^{3}\rho_{0}}{3Rr}=\frac{2GM}{\bar{r}}.
\end{eqnarray}
recalling that $M=\frac{4}{3}\pi\rho_{0}r^{3}$ and $\bar{r}=rR$ giving us the relation $k=\kappa$ .

A final consistency check is obtained now that we have derived that the internal space is a Robertson Walker space, satisfying the standard Friedmann equations, with k being interpreted as the spatial curvature, as we have just shown. Under this condition, we know the matching of this internal 
cosmology to Schwarzschild is consistent, as the analysis of the Oppenheimer- Snyder collapse model shows\cite{Weinberg}.
Here we have gone about this problem in the opposite way, showing that the matching of these two 
spaces imposes severe constraints that allows us to derive Schwarzschild space in the outside
and determine that the Newtonian constant of integration $k$ has  to be the spatial curvature of the internal FLRW internal space, all of this without using Einstein's equations.

\section{Conclusions}
In this paper it has been found that matching a dust ball, whose dynamics is governed by the Newtonian cosmology equations, containing a constant of integration $k$, to an external static space-time, where 
\begin{eqnarray}
ds^2 = - B(\bar{r}) d \bar{t}^2 + A(\bar{r})d\bar{r}^2 + \bar{r}^2d \Omega^2
\end{eqnarray}
where $B = \left( 1 - \frac{2GM}{\bar{r}}\right)$, forces $A$ to have a very special form. Assuming either asymptotic flatness, or absence of signature change, we uniquely obtain
\begin{eqnarray}
A(r) = \frac{1}{\left(1- \frac{2GM}{\bar{r}}\right)}.
\end{eqnarray}
Finally the same matching to the internal FLRW space
\begin{eqnarray}
ds^2 = -dt^2 + R^2(t)\left(\frac{dr^2}{1- \kappa r^2}+r^2 d\Omega^2\right)
\end{eqnarray}
forces the geometrical parameter $\kappa$ that appears in FLRW to coincide with the constant of integration $k$, used in the Newtonian Cosmology. 

In this essay, it has been found that matching a dust ball, whose dynamics
is governed by the Newtonian cosmology equations, containing a constant
of integration $k$ coincides with the geometrical parameter $\kappa$
that appears in FLRW used in the Newtonian Cosmology. These results
are of interest at least in two respects, one from the point of view
of its pedagogical value of teaching General Relativity without in
fact using Einstein's equation and second, the fact that some results
attributed to General Relativity can be obtained without using General
Relativity indicates that these results are more general than the
particular dynamics specified by General Relativity. Although,
some generalizations are possible. like the possibility of introducing a cosmological
constant in Newtonian Cosmology, as discussed in Bondi’s book \cite{ref4}
, we could in this way that by matching this cosmology to an exterior stationary space we obtain Schwarzschild- deSitter space, but EE do not expect that an approach of this type will be able to give all
results of GR, for example, certainly not in the case of gravitational waves or the Kerr solutions.For the case of inhomogeneous dust ball distribution, the CM lies on arbitrary origin and hence the linear and angular momentum remains zero. So also has been shown that the effective force between the constituent particles are zero for no perturbation \cite{ref5}.If we ponder minutely, the paper \cite{ref6}  goes very far just using Newtonian physics, but is missing a space-time interpretation of the solutions of Newton’s laws for gravitationally interacting particles. We might try to find them properly even in presence of pressure in our next venture.

 Finally an interesting question arises that whether our derivation holds only in the weak field approximation or not. Notice that indeed in parts of our arguments we have used
the weak field approximation, like when we derived the $0-0$ component
of the metric, then on this basis, we derived the $r-r$ component
by a process of matching to the internal collapsing ball of dust.
But if we take the point of view that we trust the metric of the collapsing
dust beyond the weak field approximation, the situation will be different,
in this case our derivation could have validity beyond the weak field
approximation, this is a question to be studied. One should also point
out that in GR, as far as the post-Newtonian approximation is concerned,
the corrections to the the $0-0$ component of the metric appear at
the same other as the first corrections to the $r-r$ components,
so it is in a sense puzzling that the $r-r$ component has been more
elusive to find by a simple derivation, as compared to the case of
the $0-0$ component.

\section{Acknowledgment}
E.G. wants to thank the Fresno California State University Physics Department for the opportunity to present a preliminary version of the results of this paper in a colloquium.


\begin{thebibliography}{99}
 \bibitem{ref1} A. Sommerfeld :"Vorlesungen \"uber theoretische Physik" vol. 3 (Leipzig, 1949), translated in "Electrodynamics (Lecture Notes on theoretical Physics, NY, Academic Press)
\bibitem{ref2} R.P. Gruber, R.H. Price, S.M. Matthews, W.R. Cordwell and L.F. Wagner: "The impossibility of a simple derivation of the Schwarzschild metric" Am. J. Phys. {\bf 56} 265 (1998), 
W. Rindler: "Counterexample to the Lenz Shift Argument", Am. J. Phys. {\bf 36} 540 (1968).

\bibitem{ref3} A. Rabinowitz:"Magically obtaining Einstein from Newton, via reverse Platonic projection", http://www.youtube.com/watch?v=Luz4VFGaXI. Lecture delivered at Ben Gurion University.
\bibitem{Visser} M. Visser:Int.J.Mod.Phys. D14 (2005) 2051,  
e-Print: gr-qc/0309072 
\bibitem{ref4} Milne, E. A.:Quart. J. Math., 5, 64 (1934), 
 McCrea, W. H. , and Milne, E. A.: Quart. J. Math., 5, 73 (1934), 
  McCrea, W. H.:"Rep. Prog. Phys.", 16, 321 (1953),
 Layzer, D.:Astro. J., 59, 268 (1954), 
 Bondi, H.:"Cosmology", 78 (Camb. Univ. Press, 1952),
 Bondi, H.:Mon. Not. Roy. Astro. Soc., 107, 410 (1947).
 \bibitem{Gautreau} R. Gautreau:Phys. Rev. D29, 186 (1984).
\bibitem{Weinberg} Weinberg, Steven. Gravitation and Cosmology. New York, NY: Wiley, 1972. ISBN: 9780471925675. Pages 342-350.
\bibitem{ref5}
George F.R.Ellis,Gary W. Gibbons:Discrete Newtonian Cosmology:Perturbations,Class.Quant.Grav. 32 (2015) 5, 055001
 arXiv:1409.0395 [gr-qc] 

\bibitem{ref6} George F.R. Ellis, Gary W. Gibbons:Discrete Newtonian Cosmology, Class.Quant.Grav. 31 (2014) 025003, arXiv:1308.1852 [astro-ph.CO] 

\bibitem{Ref7}
Counterexample to the Lenz-Schiff Argument
W. Rindler , Am. J.  Phys  36, 540 (1968); doi: 10.1119/1.1974967

\bibitem{Ref8}
Schwarzschild and de Sitter solutions from the argument by Lenz and Sommerfeld

Am. J. Phys. 79, 662 (2011); 10.1119/1.3557070

\end{thebibliography}
\end{document}